\documentclass[12pt]{elsarticle}
\usepackage{amsmath}
\usepackage{xcolor}
\usepackage{stackengine}
\usepackage{braket}
\usepackage{comment}
\usepackage{bbold}
\usepackage{subcaption}
\usepackage{graphicx}
\usepackage[a4paper, total={7.2in, 10in}]{geometry}
\title{Twin-field-based multi-party quantum key agreement}
\author{Venkat Abhignan}
 \address{Qdit Labs Pvt. Ltd., Bengaluru - 560092, India}
 \author{R. Srikanth}
 \address{Theoretical Sciences Division, Poornaprajna Institute of Scientific Research (PPISR), Bengaluru-562164, India}

\begin{document}
\makeatletter
\def\ps@pprintTitle{%
  \let\@oddhead\@empty
  \let\@evenhead\@empty
  \let\@oddfoot\@empty
  \let\@evenfoot\@oddfoot
}
\makeatother

\begin{frontmatter}
\begin{abstract}
 Quantum key distribution (QKD) can secure cryptographic communication between two distant users, as guaranteed by the laws of quantum mechanics rather than computational assumptions. The twin-field scheme, which employs counter-propagated weak coherent light pulses, doubles the secure distance of standard QKD without using quantum repeaters. Here, we study a method to extend the twin-field key distribution protocol to a scheme for multi-party quantum key agreement.   We study our protocol's security using a minimum error discrimination analysis and derive the asymptotic key rate based on the entanglement-based source-replacement scheme. We also simulate it on the ANSYS Interconnect platform with optical components to study the protocol's performance in certain practical situations. \color{black}
\end{abstract}
\end{frontmatter}
\section{Introduction}

The ability of quantum key distribution (QKD) \cite{PhysRevLett.67.661, BENNETT20147} to provide two remote users, Alice and Bob, with a secret key that is both composable and unconditionally secure with the principles of quantum mechanics is remarkable in quantum communication and cryptography \cite{RevModPhys.81.1301,Lo2014,RevModPhys.92.025002,Pirandola:20}. Recently, there has been development in practical implementations of secure quantum communication across longer distances due to novel twin-field QKD protocol \cite{Lucamarini2018}. The maximum key rate for traditional point-to-point QKD systems increases linearly with the channel transmittance $\eta$ based on the losses \cite{Takeoka2014,Pirandola2017}. Twin-field QKD offers a better rate-loss tradeoff since the key rate grows linearly with $\sqrt{\eta}$ \cite{Minder2019,Yin2019,PhysRevA.107.042603}. 

Typical QKD systems, including twin-field protocols, facilitate the exchange of keys only between two parties. Development in quantum communication over quantum networks for applications like quantum internet \cite{Castelvecchi2017} require an exchange of keys between multiple parties \cite{elliott2005current,4787043,Peev_2009,Wang:10,Stucki_2011,Sasaki:11,PhysRevX.6.011024,Chen2021}. QKD networks are typically built based on reliable central relays \cite{Peev_2009,Stucki_2011,Sasaki:11,Chen2021}, which is not ideal as any successful attack on the relay would compromise the network's security. The central relays also limit the key rates with the channel transmittance $\eta$ \cite{elliott2005current,4787043,Wang:10,PhysRevX.6.011024}.  

A twin-field QKD network would have a remarkable advantage over existing networks in terms of secure key rate, and it is intrinsically a measurement-device-independent QKD impervious to some side-channel attacks \cite{PhysRevLett.108.130502,PhysRevLett.108.130503}. A Sagnac interferometer-based twin-field QKD for multiple parties was implemented recently for pairwise communication \cite{Zhong2021,PhysRevApplied.17.014025,Park2022}. The twin-field scheme can also be applied to standard quantum cryptography protocols employing single photons or entangled states, for example, quantum digital signature \cite{Rao_2023} constructed on the basis of the counterfactual principle \cite{noh2009counterfactual, PhysRevA.109.032435}. 

\color{black}

Quantum key agreement protocols attempt to share secret keys among several parties \cite{PhysRevA.57.822,cabello2000multipartykeydistributionsecret,Chen2007,PhysRevA.76.062316}. It has been demonstrated that quantum key agreement has advantages compared to repeated QKD in quantum networks \cite{Epping_2017}. Many quantum key agreement protocols have been suggested \cite{Murta2020}, including measurement-device-independent \cite{PhysRevLett.114.090501} and device-independent \cite{PhysRevA.97.022307} versions. However, similar to QKD, $\eta$ restricts the key rates in most quantum key agreement protocols \cite{PhysRevX.11.041016}. Twin-field QKD has been implemented in quantum key agreement protocols for symmetric \cite{Grasselli_2019,PhysRevApplied.14.024010,9541200,Cao_2021} and asymmetric quantum channels \cite{Li_2021} to break this constraint. 

In this spirit, we propose a simple technique for a multi-party quantum key agreement procedure in which three parties share a key concurrently using the twin-field protocol, which can be straightforwardly extended for more parties. Extending twin-field to multi-field is not feasible as the experimental complications such as phase stabilization get compounded \cite{Minder2019,PhysRevLett.123.100505,Fang2020,PhysRevLett.124.070501}. Thus, we reduce the problem to a twin-field implementation where existing technology can be used for a higher number of parties. Moreover, this technique works for the arbitrary topology of a practical network, not just cyclic Sagnac interferometer-based twin-field \cite{Zhong2021,PhysRevApplied.17.014025,Park2022}.

The paper is sectioned as follows: The protocol is described in Sec. 2 with minimum error discrimination analysis against individual attacks in Sec. 3. The asymptotic key rate based on security against collective attacks is demonstrated in Sec. 4. The simulations with optical components for the proposed multi-party quantum key agreement procedure are performed on the ANSYS Interconnect platform and the details are given in Sec. 5. Further, some difficulties in the practical implementation of this protocol are discussed in Sec. 6. 

\section{Multi-party quantum key agreement based on twin-field cryptography}
\begin{figure}[htp]
    \centering
\includegraphics[width=8cm]{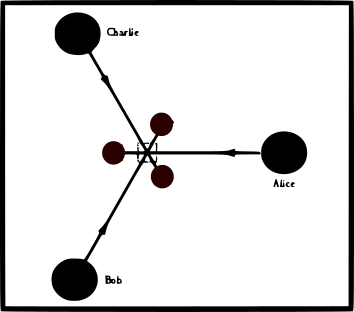}
    \caption{\color{black} Possible tri-party triple-field implementation of quantum key agreement: three-mode linear interferometer injected with three weak coherent pulses.}
    \label{fig:triple-field}
\end{figure}

\color{black}
In multiparty quantum key agreement protocols the key rate for $N$ parties generally suffers as O($\eta^N$) or O($\eta^{N-1}$) \cite{PhysRevApplied.14.024010} and extending the twin-field implementation for $N>3$ parties is non-trivial \cite{9541200,Cao_2021,Li_2021}. Current technology does not facilitate the experimental demonstration of the scheme implemented in Ref. \cite{Grasselli_2019}. Even considering a triple-field interference (where Alice, Bob, and Charlie send pulses to a central hub) such as in Fig. \ref{fig:triple-field} would create large misalignment issues, and because the combinatorics of participants' choice grows exponentially with N, it would generally not be favorable.

Thus, we are led to consider the problem of extending the twin-field scheme to the multiparty case so that the standard twin-field protocol remains a fixed subroutine applied by different sets of parties spread out over a wide area. Our simple but fundamental observation here is that key agreement (as against key distribution) may circumvent the distance factor because it involves multiple trusted parties, making it likely that any given party will always find another involved in the close neighborhood. Our protocol exploits this property to optimize the quantum resources used. We begin by considering a three-party protocol. \color{black}

\subsection{Description of the three-party protocol}

\begin{figure}[htp]
    \centering
\includegraphics[width=17cm]{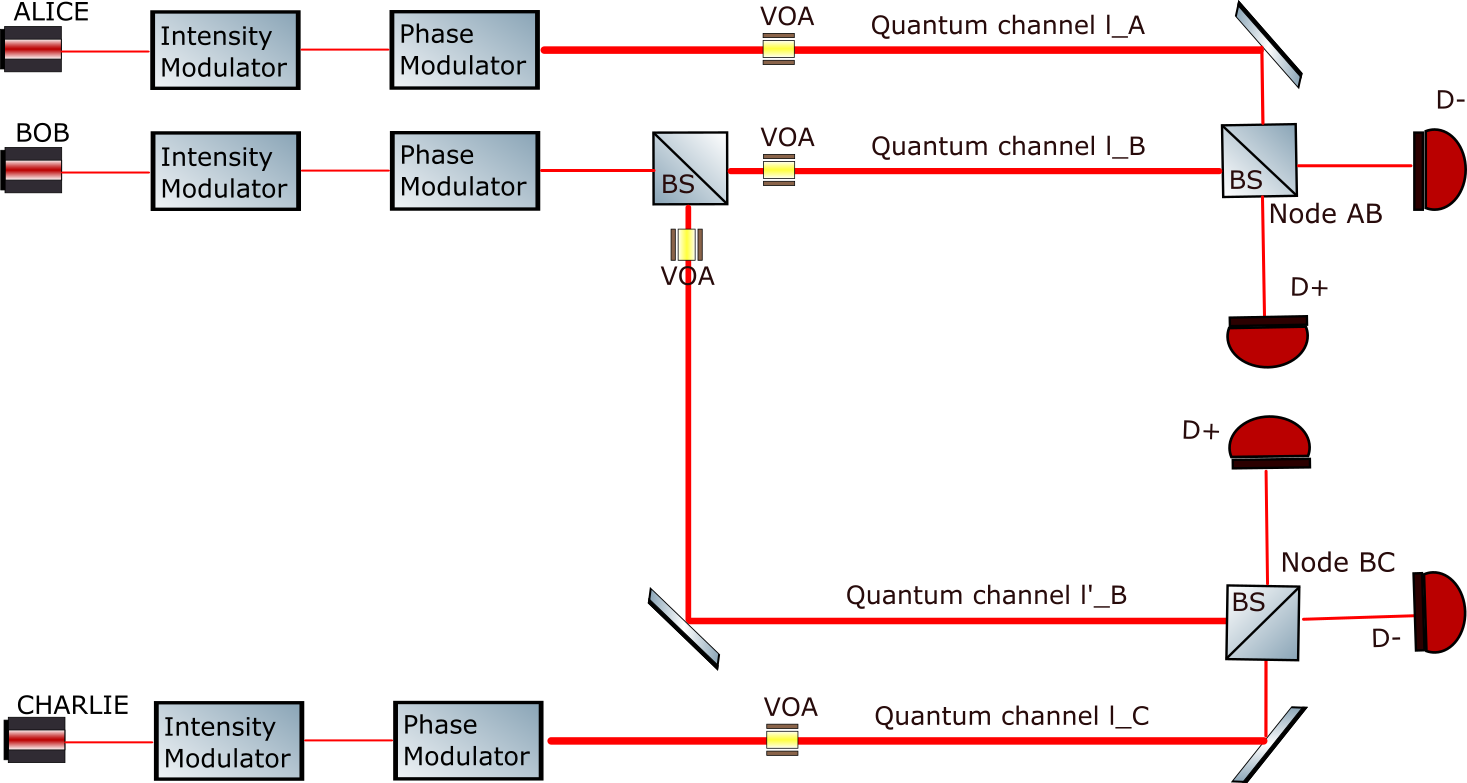}
    \caption{Schematic setup of multi-party twin-field QKD protocol. Alice and Bob send their coherent states to node AB, whereas Bob and Charlie send coherent states to node BC. BS: 50-50 beam splitter. D+,D-: single-photon detectors. VOA: variable optical attenuator.}
    \label{multiparty}
\end{figure}

In the three-party twin-field protocol, Alice, Bob, and Charlie send information to two nodes, AB and BC, as shown in Fig. \ref{multiparty}. For $N$-party protocol, $N-1$ nodes are required as per this conception.
\begin{enumerate}
    \item Alice, Bob and Charlie choose a coherent state $\ket{\alpha\text{e}^{\text{i}\pi k_A}}$, $\ket{\beta\text{e}^{\text{i}\pi k_B}}$ and $\ket{\gamma\text{e}^{\text{i}\pi k_C}}$ and a bit value $k_A,k_B,k_C$  $\in \{0, 1\}$ with equal probability. Alice sends it through the quantum channels $l_A$, Bob sends it through $l_B$ and $l_B'$ (By placing a fibre optic splitter before the variable optical attenuator at Bob's side, she can produce two channels. The classical signal is split into two and then two quantum signals are obtained.), and Charlie sends it through $l_C$. Channel $l_A$ and $l_B$ link to an intermediary node AB, and $l_B'$ and $l_C$ link to intermediary node BC. They select the timings so that the signals reach the nodes simultaneously. Additionally, they calibrate the amplitudes of their coherent states $\alpha,\beta$ and $\gamma$ so that the states arrive at the node with the same amplitude after passing via the channels of length $l_A$, $l_B$, $l_B'$ and $l_C$.
    \item When two signals arrive at the intermediate nodes AB and BC if the node AB(BC) finds a pair of states that are correlated, it declares $\delta_1(\delta_2)=``+(+)"$; if it finds an anti-correlated pair, it announces $\delta_1(\delta_2)=``-(-)"$, and if the result is unclear, it announces $\delta_1(\delta_2)=``?(?)"$ ($+$ corresponds to click in detector $D+$, $-$ corresponds to click in detector $D-$). The subsets of data that Alice and Bob gathered are separated and indexed by $\delta_1$, while data gathered by Bob and Charlie are indexed by $\delta_2$. 
    \item Following the announcement from nodes AB(BC), Alice, Bob and Charlie post-process their data as follows:
    \begin{enumerate} 
    \item At AB node, if $\delta_1 = +$, Alice and Bob store their respective outcomes as the shared secret bit. If $\delta_1 = -$, Alice flips her bit value. Repeating this, Alice and Bob obtain key $k_{AB}$.
    \item If $\delta_2 = +$ at the BC node, Bob and Charlie store their respective outcomes as the shared bits. If $\delta_2 = -$, Charlie flips his bit value. Repeating this, Charlie and Bob obtain key $k_{BC}$.
    \item \textcolor{black}{Suppose $|k_{AB}|=|k_{BC}|$. Bob publicly announces $k_{AB+BC} \equiv k_{AB} \oplus k_{BC}$. Using her (resp., his) private copy of $k_{AB}$ (resp., $k_{BC})$, Alice (resp., Charlie) retrieves $k_{AB} = k_{BC} \oplus k_{AB+BC}$ (resp., $k_{BC} = k_{AB} \oplus k_{AB+BC}$). If $|k_{AB}|\ne|k_{BC}|$, then $||k_{AB}|-|k_{BC}||$ bits are removed by public discussion from the longer key, and the above procedure is applied.}
    \end{enumerate}
    \item To derive the final joint secret key, Alice, Bob, and Charlie subject the consolidated key $K$ (based on $|k_{AB}|$ and $|k_{BC}|$) to privacy amplification and error correction.
    \color{black}
\end{enumerate}

\begin{figure}[htp]
    \centering
\includegraphics[width=8cm]{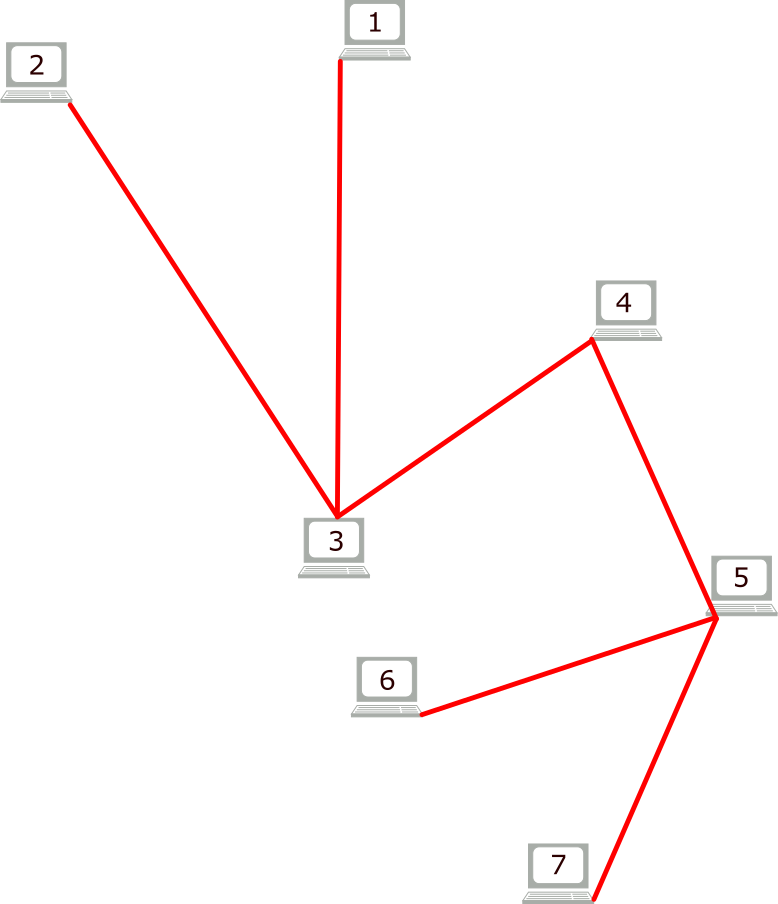}
    \caption{\color{black} Network with $N=7$ parties and $n=3$ twin-field key agreement implementations, among the parties $\{1,3,2\}, \{3,4,5\}$ and $\{6,5,7\}$. Public announcements are made at vertices 3, 4 and 5. The key rate does not scale linearly with the size of the network (identified with the maximum distance between two parties, in this case, $d_{1,7}$), but instead quadratically between the longest nearest-neighbor distance, in this case, $d_{2,3}$.}
    \label{fig:network}
\end{figure}

\color{black}
To generalize the above protocol, suppose we are given a group of $N>3$ parties that must agree on a key. First, a classical optimization algorithm is run to determine the optimal network that links all parties while minimizing the total length of all edges. Barring material or cost obstacles (which would be additional constraints, if applicable), this optimal network will constitute the required quantum network. The network's vertices are divided into $n$ segments of three connected vertices, so each segment shares a vertex with at least one other segment. Ideally, we have $2n+1=\lceil N\rceil$, where the ceiling function represents round off to the following highest odd number (cf. Fig. \ref{fig:network}). 

Each segment will run the twin-field quantum key agreement protocol described above. The keys obtained in individual segments can be reconciled by \textit{inter-segmental} public announcements by the parties shared between segments (different from but analogous to the \textit{intra-segmental} announcements mentioned above). Suppose the keys (assumed to be of equal size) obtained from two contiguous twin-field segments are $s_1$ (for instance, $K$ in the above description) and $s_2$. Then, the XOR of the two keys, $s_{12} \equiv s_1 \oplus s_2$, is announced. The remaining members of the first segment can derive $s_2 = s_{12}\oplus s_1$, similarly for the other segment. The number $m$ of inter-segmental announcing parties generally depends on the network topology. For the circuit in Fig. \ref{fig:network}, $m=3$, whereas for a network with star topology (a hub vertex, with the rest radiating from it), $m=1$ corresponds to the announcing party at the hub.

A practical consequence of our approach is that the key rate in our protocol will scale as O($\sqrt{\eta}$) of the longest path between two nearest neighbors in this optimal circuit, irrespective of the geographic spread of the $N$ parties. For example, in Fig. \ref{fig:network}, the key rate scales quadratically as $d_{2,3}$ ($\eta\propto d$) rather than as the longest distance between two parties $d_{1,7}$. 
\color{black}

\color{black}

\subsection{Security aspects of multi-party quantum key agreement}

When concerned with the security of quantum key agreement protocols, Eve, an eavesdropper, is the outside entity that can listen to all public communications. Eve may also attempt to manipulate the quantum channels and investigate correlations with the exchanged information. Theoretically, the Eve's attacks are grouped into individual, collective, and coherent attacks \cite{Murta2020}. Each protocol round can only be attacked by Eve separately in individual attacks. Her best action in this situation is to measure her quantum side information at each round, as it is presumed that she lacks quantum memory. We perform minimum error discrimination analysis \cite{Ferenczi} in Sec. 3, which can be considered an intercept-resend strategy for individual attacks. In collective attacks, Eve is thought to carry out the same attack for every protocol round, meaning that the quantum side information is independent and identically distributed (IID) across rounds. It is assumed here that Eve has a quantum memory, and as a result, she can execute a global operation at the end of the protocol and save her quantum side information for each round. We perform this analysis in Sec. 4 using source-replacement strategy \cite{PhysRevLett.92.217903,PhysRevA.98.042332} to obtain the asymptotic secret key rate. Coherent attacks are the most general attack in which Eve's capabilities are not limited, and they are generally used to study finite-size effects in the key. 

\color{black}

\section{\color{black} Individual attacks and minimum error discrimination analysis}
Initially we perform the simple minimum error discrimination analysis (on individual pulses) on the multi-party protocol to show its robustness compared to two-party protocol. Assuming that Eve controls the nodes AB, BC and the channels $l_A$, $l_B$, $l'_B$ and $l_C$, she makes the announcements $\delta_1(\delta_2)$ \cite{Ferenczi,PhysRevA.98.042332}. Eve must effectively discriminate between the joint states that are correlated vs anti-correlated at node AB
\begin{equation}
    \rho^{AB}_{\pm}=\frac{1}{2}(\ket{\alpha,\pm\beta}\bra{\alpha,\pm\beta}+\ket{-\alpha,\mp\beta}\bra{-\alpha,\mp\beta}),
\end{equation}
and likewise at node BC,
\begin{equation}
\rho^{BC}_{\pm}=\frac{1}{2}(\ket{\beta,\pm\gamma}\bra{\beta,\pm\gamma} + \ket{-\beta,\mp\gamma}\bra{-\beta,\mp\gamma}).
\end{equation}
Considering Eve does not make any inconclusive announcements with ``$\delta_1(\delta_2)=?(\pm),$ $\pm(?), ?(?)$", she only makes announcements with $\delta_1(\delta_2)=\pm(\pm)$. Due to their non-orthogonality of the joint states, it is impossible to identify the two states $\rho^{\text{AB(BC)}}_+$ and $\rho^{\text{AB(BC)}}_-$ without making mistakes; and Eve occasionally announces something incorrectly. For example, she might announce $-(+)$ while the state was actually $\rho^{\text{AB(BC)}}_+(\rho^{\text{AB(BC)}}_-)$. The minimum error discrimination strategy gives the minimum error to discriminate between correlated and anti-correlated states as \cite{Helstrom1969} 
 \begin{equation}
Q_{\text{min}}^{\text{ABC}}= 
1-(1-Q_{\text{min}}^{\text{AB}})(1-Q_{\text{min}}^{\text{BC}})
\end{equation} 
where \begin{equation*}
    Q_{\text{min}}^{\text{AB(BC)}}= Q_{\text{min}}^{\text{A(B)}}(1-Q_{\text{min}}^{\text{B(C)}}) +
    Q_{\text{min}}^{\text{B(C)}}(1-Q_{\text{min}}^{\text{A(B)}}).
\end{equation*}
It can be shown that
\begin{equation} Q_{\text{min}}^{\text{AB(BC)}}=\frac{1}{2}\left(1-|\frac{1}{2}(\rho^{\text{AB(BC)}}_--\rho^{\text{AB(BC)}}_+)|_1\right).
 \end{equation}
Further, the typical infinite-dimensional Fock space description of states $\pm\ket{\alpha}$ can span a two-dimensional subspace such as \begin{equation}
        \ket{\pm\alpha} = c_0\ket{e_0}\pm c_1\ket{e_1},
\end{equation} with normalized $|c_0|^2+|c_1|^2=1$ and orthonormal basis $\{\ket{e_0},\ket{e_1}\}$. This canonical basis can be written in Fock state basis as \cite{PhysRevA.98.042332} \begin{subequations}
    \begin{align}
        \ket{e_0}&=\frac{1}{\sqrt{\text{cosh}(|\alpha|^2)}}\sum_{n=0}^{\infty}\frac{\alpha^{2n}}{(2n)!}\ket{2n},\\
        \ket{e_1}&=\frac{1}{\sqrt{\text{sinh}(|\alpha|^2)}}\sum_{n=0}^{\infty}\frac{\alpha^{2n+1}}{(2n+1)!}\ket{2n+1},
    \end{align}
\end{subequations}
with $c_0=\text{e}^{-|\alpha|^2/2}\sqrt{\text{cosh}(|\alpha|^2)}$ and $c_1=\text{e}^{-|\alpha|^2/2}\sqrt{\text{sinh}(|\alpha|^2)}$.  We can obtain similar expressions for states $\{\pm\ket{\beta}\},\{\pm\ket{\gamma}\}$.
Using the basis elements in a space of $\{\ket{e_0,e_0},\ket{e_1,e_1},\ket{e_0,e_1},\ket{e_1,e_0}\}$ in Eq. (1) and Eq. (2) we can obtain (for convenience we set $|\alpha|^2=|\beta|^2=|\gamma|^2 = \mu$, same intensity for the coherent states) 
\begin{align}
 \rho^{AB(BC)}_+ = \begin{pmatrix}
    2 c_0^4 & -2 c_0^2 c_1^2 & 0 & 0 \\
-2 c_0^2 c_1^2 & 2 c_1^4 & 0 & 0  \\
0 & 0 & 2 c_0^2 c_1^2 & -2 c_0^2 c_1^2  \\
0 & 0 & -2 c_0^2 c_1^2 & 2 c_0^2 c_1^2
  \end{pmatrix},
\end{align}
\begin{align}
 \rho^{AB(BC)}_- = \begin{pmatrix}
    2 c_0^4 & 2 c_0^2 c_1^2 & 0 & 0 \\
2 c_0^2 c_1^2 & 2 c_1^4 & 0 & 0  \\
0 & 0 & 2 c_0^2 c_1^2 & 2 c_0^2 c_1^2  \\
0 & 0 & 2 c_0^2 c_1^2 & 2 c_0^2 c_1^2
  \end{pmatrix}.
\end{align}
Implementing this in Eqs. (4) and (3), we can obtain the minimum error as,
\begin{subequations}
    \begin{align}   Q_{\text{min}}^{\text{AB(BC)}}&=\frac{1}{2}(1-\text{e}^{-2\mu}\text{cosh}(\mu)\text{sinh}(\mu))=\frac{1}{8} \left(e^{-4 \mu}+3\right),\\
    Q_{\text{min}}^{\text{ABC}}&=1-\left(\frac{1}{2} \left(e^{-2 \mu} \sinh (\mu) \cosh (\mu)-1\right)+1\right)^2=\frac{1}{64} \left(-e^{-8 \mu}+10 e^{-4 \mu}+39\right).
    \end{align}
\end{subequations}
This shows that minimum error for Eve to discriminate between $\rho^{\text{AB(BC)}}_+$ and $\rho^{\text{AB(BC)}}_-$ is always more for three parties $Q_{\text{min}}^{\text{ABC}}$ than for two parties $Q_{\text{min}}^{\text{AB(BC)}}$ for $\mu<1$, where the optimal intensity  usually is $\mu_{\text{opt}}<1$ for high key rates \cite{PhysRevA.98.042332}. 

In a iterative manner, we can obtain minimum error for Eve to discriminate between non-orthogonal states at three independent nodes between four parties which will be more than for three parties.

\section{\color{black}Collective attacks and the asymptotic key rate}
To demonstrate the security against collective attacks and asymptotic key rate of the twin-field protocol, Ref. \cite{PhysRevA.98.042332} introduces the source-replacement strategy \cite{PhysRevLett.92.217903}, which is applied  to Alice and Bob's sources. Here we extend this to Charlie's source. Then, we adapt the multi-party protocol to its equivalent entanglement-based protocol to verify the security against collective attacks in the asymptotic-key scenario. The key rate is then assessed to demonstrate the entanglement-based version's security. 

In the source-replacement scheme for every round, Alice, Bob and Charlie select signal states $\{\ket{\varphi_x},\ket{\varphi_y},\ket{\varphi_z}\}$ from an a priori probability distributions $\{p_x,q_y,r_z\}$, respectively. The state $\ket{\psi}_{ABA'B'}$ between Alice-Bob in the source-replacement scheme is \begin{equation}
\ket{\psi}_{ABA'B'} =\textstyle\sum_{x,y} \sqrt{p_x q_y} \ket{x,y}_{AB}\ket{\varphi_x,\varphi_y}_{A'B'} 
\end{equation} 
and the state $\ket{\psi}_{BCB'C'}$ between Bob-Charlie is 
\begin{equation}
\ket{\psi}_{BCB'C'} =\textstyle\sum_{y,z} \sqrt{q_y r_z} \ket{y,z}_{BC}\ket{\varphi_y,\varphi_z}_{B'C'}. 
\end{equation} 
Here, Alice, Bob, Charlie record the choices of states prepared in registers $A$, $B$, $C$ based on orthonormal basis $\{\ket{x},\ket{y},\ket{z}\}$ and in registers $A'$, $B'$, $C'$ corresponding to states $\{\ket{\varphi_x},\ket{\varphi_y},\ket{\varphi_z}\}$, respectively. While information in $A$, $B$, $C$ of Alice, Bob, and Charlie are kept secret from Eve where they perform a local operation on their registers to measure outcome, states $\ket{\varphi_x,\varphi_y}_{A'B'}$ and $\ket{\varphi_y,\varphi_z}_{B'C'}$ in registers $A'$, $B'$, $C'$ are sent to nodes AB and BC.

 Eve independently takes measurements characterised by a POVM $F^{\delta_1(\delta_2)}$ on the states at nodes AB and BC. States from Alice, Bob and Charlie in the registers $A'$, $B'$ and $C'$ are measured to create a strategy for her announcements $\delta_1$ and $\delta_2$. $F^{\delta_1(\delta_2)}$ is defined for $F^{\delta_1}=F^{\delta_2}={F^+,F^-,F^?}$, where outcomes $\delta_1,\delta_2 \in \{+,-,?\}$ are relevant for Alice, Bob and Charlie to reconcile a key as can be seen in Sec. 2.1. Eve applies on the input quantum states $\ket{\varphi_x,\varphi_y}_{A'B'}$ and $\ket{\varphi_y,\varphi_z}_{B'C'}$ in registers $A'$, $B'$, $C'$ a completely positive trace-preserving (CPTP) map. The announcements regarding the measurements of states $\ket{\varphi_x,\varphi_y}_{A'B'}$ and $\ket{\varphi_y,\varphi_z}_{B'C'}$ are recorded in the classical registers $C_1$ and $C_2$ corresponding to $\delta_1$ and $\delta_2$, while the post-measurement state is maintained in the registers E and F, with an orthonormal basis $\{\ket{\delta_1}\}$ and $\{\ket{\delta_2}\}$ to obtain joint states $\rho_{ABEC_1}$ at node AB and $\rho_{BCFC_2}$ at node BC, respectively. 
 
Alice, Bob and Charlie use POVMs on the $A$, $B$ and $C$ registers to perform measurements. Alice, Bob and Charlie keep their measurement results in classical $X$, $L$ and $M$ registers, respectively. Alice then uses a CPTP map to translate the measurements in register $X$ to the raw key bit in register $K$, while Bob uses a CPTP map to translate the measurements in register $L$ to the raw key bit in register $K'$. With this, the joint states corresponding to node AB and BC become $\rho_{ABEC_1} \rightarrow \rho_{KLEC_1}$ and $\rho_{BCFC_2} \rightarrow \rho_{K'MFC_2}$. Using $\rho_{KLEC_1}$ and $\rho_{K'MFC_2}$, conditional states for Eve $\rho^{k,l,\delta_1}_{E}$and $\rho^{k',m,\delta_1}_{F}$ are defined such that Alice-Bob hold $k,l$ in registers $K,L$ and Bob-Charlie hold $k^{\prime},m$ in registers $K',M$. 
  
  $\rho^{k,l\delta_1}_{E}$ and $\rho^{k',m,\delta_2}_{F}$ are defined as \begin{equation}
\rho^{k,l,\delta_1}_{E} = \ket{\Theta^{\delta_1}_{k,l}}\bra{\Theta^{\delta_1}_{k,l}} \ \text{and} \
\rho^{k',m,\delta_2}_{F} = \ket{\Theta^{\delta_2}_{k',m}}\bra{\Theta^{\delta_2}_{k',m}}
\end{equation}
 where 
 \begin{equation}
\ket{\Theta^{\delta_1}_{k,l}} = \frac{\sqrt{F^{\delta_1}}\ket{\varphi_k,\varphi_l}}{\sqrt{\bra{\varphi_k,\varphi_l}F^{\delta_1}\ket{\varphi_k,\varphi_l}}},\ket{\Theta^{\delta_2}_{k',m}} = \frac{\sqrt{F^{\delta_2}}\ket{\varphi_{k'},\varphi_m}}{\sqrt{\bra{\varphi_{k'},\varphi_m}F^{\delta_2}\ket{\varphi_{k'},\varphi_m}}}
 \end{equation}
 since POVM elements $F^{\delta_2}$ and $F^{\delta_1}$ are independently used for measurement at nodes AB and BC by Eve.
 
 Further, states $\rho^{k,\delta_1}_{E}$,$\rho^{\delta_1}_{E}$ at node AB and $\rho^{k',\delta_2}_{F}$,$\rho^{\delta_2}_{F}$  at node BC are defined as 
\begin{equation}
\rho^{k,\delta_1}_{E}=\sum_{l}p(l|k,\delta_1)\rho^{k,l,\delta_1}_{E}, 
\rho^{\delta_1}_{E}=\sum_{k}p(k|\delta_1)\rho^{k,\delta_1}_{E}
    \end{equation}
and 
\begin{equation}
\rho^{k',\delta_2}_{F}=\sum_{m}p(m|k',\delta_2)\rho^{k',m,\delta_2}_{F},
\rho^{\delta_2}_{F}=\sum_{k'}p(k'|\delta_2)\rho^{k',\delta_2}_{F}.
    \end{equation}

The total number of secret bits that can be extracted from the state $\rho_{KLEC_1}$ at node AB and state $\rho_{K'MFC_2}$ at node BC is defined as 
\begin{equation}
\Tilde{r}(\rho_{KLEC_1})=\sum_{\delta_1} p(\delta_1) r(\rho^{\delta_1}_{KLE}) \ \hbox{and} \ \Tilde{r}(\rho_{K'MFC_2})=\sum_{\delta_2} p(\delta_2) r(\rho^{\delta_2}_{K'MF}).
\end{equation} Based on this, we define the total number of secret bits that can be extracted from the states $\rho_{KLEC_1}$ at node AB and $\rho_{K'MFC_2}$ at node BC as 
\begin{equation}
\Tilde{r}(\rho_{KLEC_1},\rho_{K'MFC_2})= \min\bigg(\Tilde{r}(\rho_{KLEC_1}), \Tilde{r}(\rho_{K'MFC_2})\bigg).
\end{equation} 
Consecutively, the number of secret bits that we can extract from the states $\rho^{\delta_1}_{KLE}$ at node AB and $\rho^{\delta_2}_{K'MF}$ at node BC based on Devetak-Winter formula \cite{Devetak} are defined as \begin{equation}
r(\rho^{\delta_1}_{KLE})=\text{max}[1-\delta^{\delta_1}_{\text{EC}}-\chi(K:E)_{\rho^{\delta_1}_{KLE}},0] \ \text{and} \ 
r(\rho^{\delta_2}_{K'MF})=\text{max}[1-\delta^{\delta_2}_{\text{EC}}-\chi(K':F)_{\rho^{\delta_2}_{K'MF}},0].
\end{equation} Here $\delta^{\delta_1(\delta_2)}_{\text{EC}}$ is the amount of information leakage per signal during the error correcting phase, and Holevo information is
\begin{equation}
\chi(K:E)_{\rho^{\delta_1}_{KLE}}=S(\rho^{\delta_1}_{E})-\sum_{k} p(k|\delta_1)S(\rho^{k,\delta_1}_{E}) \ \text{and} \ \chi(K':F)_{\rho^{\delta_2}_{K'MF}}=S(\rho^{\delta_2}_{F})-\sum_{k'} p(k'|\delta_2)S(\rho^{k',\delta_2}_{F})
\end{equation}
with $S(\rho)$ being the von Neumann entropy.  
Further, using these the asymptotic key rate $R^\infty$ is defined as \begin{equation}
R^\infty=\underset{\rho_{KLEC_1}\in C1',\rho_{K'MFC_2}\in C2'}{\min}\Tilde{r}(\rho_{KLEC_1},\rho_{K'MFC_2})
\end{equation} from Eq. (17), where $C1',C2'\in\{\rho_{KLEC_1},\rho_{K'MFC_2}:\rho_{KLEC_1}\rho_{K'MFC_2}=G'(\rho_{ABEC_1},\rho_{BCFC_2}),$ with $\rho_{ABEC_1},\rho_{BCFC_2}\in E',F'\}$ and $E',F' = \{\rho_{ABEC_1},\rho_{BCFC_2}:\rho_{ABEC_1},\rho_{BCFC_2}$ is compatible with experimental observations\}.

We consider the loss-only situation \cite{PhysRevA.98.042332}  and demonstrate the closed form expression for key rate of this protocol with a single-photon transmissivity of $\eta_1$ between Alice-Bob with $\mu_1$ intensity of coherent states at node AB, and transmissivity $\eta_2$ between Bob-Charlie with $\mu_2$ intensity states at node BC. To ensure that the observed statistics during the parameter estimation step at Alice-Bob and Bob-Charlier are consistent with Eve doing the measurements at nodes AB and BC, POVM elements $F^{\delta_1(\delta_2)}$ are defined in a space of $\{\ket{e_0,e_0},\ket{e_1,e_1},\ket{e_0,e_1},\ket{e_1,e_0}\}$ such as \begin{equation}
\begin{aligned}
F_{\text{loss}}^{+} &= (1-\xi^2) \begin{pmatrix}
\frac{1- \xi^2 \Omega^2}{8c_0^4} & \frac{1- \xi^2 \Omega^2}{8c_0^2 c_1^2} & 0 & 0\\
\frac{1- \xi^2 \Omega^2}{8c_0^2 c_1^2} & \frac{1- \xi^2 \Omega^2}{8c_1^4} & 0 & 0\\
0&0 &\frac{1+\xi^2 \Omega^2}{8c_0^2 c_1^2} & \frac{1+ \xi^2 \Omega^2}{8c_0^2 c_1^2} \\
0&0 &\frac{1+ \xi^2 \Omega^2}{8c_0^2 c_1^2} & \frac{1+\xi^2 \Omega^2}{8c_0^2 c_1^2} 
\end{pmatrix},\\
F_{\text{loss}}^{-} &= (1-\xi^2) \begin{pmatrix}
\frac{1- \xi^2 \Omega^2}{8c_0^4} & \frac{-1+\xi^2 \Omega^2}{8c_0^2 c_1^2} & 0 & 0\\
\frac{ -1+\xi^2 \Omega^2}{8c_0^2 c_1^2} & \frac{1- \xi^2 \Omega^2}{8c_1^4} & 0 & 0\\
0&0 &\frac{1+\xi^2 \Omega^2}{8c_0^2 c_1^2} & \frac{-1-\xi^2 \Omega^2}{8c_0^2 c_1^2} \\
0&0 &\frac{-1- \xi^2 \Omega^2}{8c_0^2 c_1^2} & \frac{1+\xi^2 \Omega^2}{8c_0^2 c_1^2} 
\end{pmatrix}, \\
F_{\text{loss}}^{?} &= \xi^2 \begin{pmatrix}
\frac{(1+\Omega)^2}{4c_0^4} &0 & 0 & 0\\
0 & \frac{(1-\Omega)^2}{4c_1^4} & 0 & 0\\
0&0 &\frac{1-\Omega^2}{4c_0^2 c_1^2} &0 \\
0&0 &0 & \frac{1-\Omega^2}{4c_0^2 c_1^2} 
\end{pmatrix}
\end{aligned}
\end{equation}
where $\Omega = e^{-2(1-\sqrt{\eta})\mu}$, and $\xi = e^{- \sqrt{\eta} \mu}$ based on coherent states in Eq. (5). These only depend on transmissivity $\eta$ and intensity of coherent state $\mu$ for the loss-only scenario. The conditional probabilities 
\begin{equation}
p(\delta_1|\varphi_k,\varphi_l)=\bra{\varphi_k,\varphi_l}F^{\delta_1}\ket{\varphi_k,\varphi_l} \ , \ p(\delta_2|\varphi_{k'},\varphi_m)=\bra{\varphi_{k'},\varphi_m}F^{\delta_2}\ket{\varphi_{k'},\varphi_m}
\end{equation} at nodes AB and BC for each outcome $\delta_1,\delta_2=\pm,?$ can be computed based on the POVM elements. Eve performs the measurements on signal states from $\mathcal{S}$ for $\varphi_{k(l)}=\pm\sqrt{\mu}$ and $\varphi_{k'(m)}=\pm\sqrt{\mu}$ in Eq. (13),  \begin{equation}
\begin{aligned}
\ket{+\sqrt{\mu},+\sqrt{\mu}}&=\begin{pmatrix}c_0^2\\c_1^2\\c_0c_1\\c_0c_1\end{pmatrix}, &
\ket{-\sqrt{\mu},-\sqrt{\mu}}&=\begin{pmatrix}c_0^2\\c_1^2\\-c_0c_1\\-c_0c_1\end{pmatrix}, \\
\ket{+\sqrt{\mu},-\sqrt{\mu}}&=\begin{pmatrix}c_0^2\\-c_1^2\\-c_0c_1\\c_0c_1\end{pmatrix}, &
\ket{-\sqrt{\mu},+\sqrt{\mu}}&=\begin{pmatrix}c_0^2\\-c_1^2\\c_0c_1\\-c_0c_1\end{pmatrix}. \\
\end{aligned}
\end{equation}
The conditional probabilities at node AB and BC for these signal states are computed in Table 1.
\begin{table}\small 
\begin{center}
\caption{Conditional probability distribution of announcement outcomes $\delta_{1(2)}=\pm,?$ given the states $\ket{\varphi_{k},\varphi_{l}}$ at node AB, $\ket{\varphi_{k'},\varphi_{m}}$ at node BC in the loss-only scenario.}
\begin{tabular}{ |c|c|c|c|c| }

 \hline
$\varphi_{k(l)}, \varphi_{k'(m)}$& $ +\sqrt{\mu},+ \sqrt{\mu}$ & $ -\sqrt{\mu},  -\sqrt{\mu}$ & $+\sqrt{\mu}, -\sqrt{\mu}$ & $ -\sqrt{\mu}, + \sqrt{\mu}$ \\
\hline
$p(+|\varphi_{k(l)},\varphi_{k'(m)})$ & $1-e^{-2\sqrt{\eta}\mu}$& $1-e^{-2\sqrt{\eta}\mu}$ & 0 &0 \\

 $p(-|\varphi_{k(l)},\varphi_{k'(m)})$ & 0& 0& $1-e^{-2\sqrt{\eta}\mu}$ &$1-e^{-2\sqrt{\eta}\mu}$ \\

 $p(?|\varphi_{k(l)},\varphi_{k'(m)})$ & $e^{-2\sqrt{\eta}\mu}$ & $e^{-2\sqrt{\eta}\mu}$& $e^{-2\sqrt{\eta}\mu}$ &$e^{-2\sqrt{\eta}\mu}$ \\

 \hline
\end{tabular}
\end{center}
\end{table}
Keys from the announcements $\delta_{1}=\delta_{2}=?$ cannot be extracted and error correction is not required in the loss-only scenario, implying $\delta_{\text{EC}}^{+} = \delta_{\text{EC}}^{-}=0$ in Eq. (18). Further, to find the number of determinable secret bits in Eq. (17) we need to evaluate $\chi(K:E)$ defined in Eq. (19). We first evaluate the conditional states $\rho_{E}^{k,l,+}$, $\rho_{E}^{k,l,-}$ at node AB and $\rho_{F}^{k',m,+}$, $\rho_{F}^{k',m,-}$ at node BC from Eqs. (14) and (15). Under the loss-only scenario, whenever Alice-Bob, Bob-Charlie produce coherent states with a phase difference $\pi$, announcement at nodes AB, BC will never be $\delta_1=+,\delta_2=+$, respectively. Whenever they prepare coherent states with the same phase, announcement at nodes AB, BC will never be $\delta_1=-,\delta_2=-$, respectively. We obtain conditional probabilities for $p(l|k,\delta_1)$ at node AB and $p(m|k',\delta_2)$ at node BC in Eqs. (14) and (15) as $p(0|1,+)=0$, $p(1|0,+)=0$, $p(0|0,-)=0$, $p(1|1,-)=0$. Using this, we can obtain that $\rho^{k,\delta_1}_{E}$  and $\rho^{k',\delta_2}_{F}$ are always a pure state for $\delta_{1(2)}=\pm$ and $k\in\{0,1\}$ implying $S(\rho^{k,\delta_1}_{E})=S(\rho^{k',\delta_2}_{F})=0$ in the expression for Holevo information from Eq. (19). For evaluating the Holevo information from $S(\rho^{\delta_1}_{E})$ and $S(\rho^{\delta_2}_{F})$ in Eq. (19) we use Eq. (13) to obtain \begin{subequations}
    \begin{align}
\rho^{+}_{E} = \rho^{+}_{F} = \frac{1}{2}(\ket{\Theta^{+}_{0,0}}\bra{\Theta^{+}_{0,0}}+\ket{\Theta^{+}_{1,1}}\bra{\Theta^{+}_{1,1}}),\\
\rho^{-}_{E} = \rho^{-}_{F} =\frac{1}{2}(\ket{\Theta^{-}_{0,1}}\bra{\Theta^{-}_{0,1}}+\ket{\Theta^{-}_{1,0}}\bra{\Theta^{-}_{1,0}}).
    \end{align}
\end{subequations}
From the eigenvalues of of $\rho^{+(-)}_{E}$, $\rho^{+(-)}_{F}$, we can deduce the von Neumann entropies $S(\rho^{+(-)}_{E})$, $S(\rho^{+(-)}_{F})$ using $S(\rho)=-\text{Tr}(\rho \log_2{\rho})$. Further using them in Eq. (18), we can deduce that \begin{subequations}
    \begin{align}
r(\rho^{\delta_1}_{KLE})=1-S(\rho^{+(-)}_{E})=1-h\left(\frac{1-\text{e}^{-4\mu_1(1-\sqrt{\eta_1})}\text{e}^{-2\mu_1\sqrt{\eta_1}}}{2}\right)\\r(\rho^{\delta_2}_{K'MF})=1-S(\rho^{+(-)}_{F})=1-h\left(\frac{1-\text{e}^{-4\mu_2(1-\sqrt{\eta_2})}\text{e}^{-2\mu_2\sqrt{\eta_2}}}{2}\right)
      \end{align}
\end{subequations} from Devetak-Winter formula with binary Shannon entropy $h(z)=-z\log_2{z}-(1-z)\log_2(1-z)$. Using the rates from $r(\rho^{\delta_1}_{KLE})$, $r(\rho^{\delta_2}_{K'MF})$ and conditional probabilities in Table 1, the total number of secret bits that can be extracted can be computed from Eq. (17). With this the asymptotic key generation rate is \begin{multline} 
    R^\infty = \min\left((1-e^{-2\sqrt{\eta_1}\mu_1})\left(1-h\left(\frac{1-\text{e}^{-4\mu_1(1-\sqrt{\eta_1})}\text{e}^{-2\mu_1\sqrt{\eta_1}}}{2}\right)\right)\right.,\\\left.(1-e^{-2\sqrt{\eta_2}\mu_2})\left(1-h\left(\frac{1-\text{e}^{-4\mu_2(1-\sqrt{\eta_2})}\text{e}^{-2\mu_2\sqrt{\eta_2}}}{2}\right)\right)\right). 
\end{multline}
Considering small $\mu_1,\mu_2$, gives $R^\infty = \min(\mathcal{O}(\sqrt{\eta_1}),$ $\mathcal{O}(\sqrt{\eta_2}))$. 

\section{\color{black}ANSYS simulations and finite key rate}

\begin{figure}[htp]
    \centering
\includegraphics[width=25cm, angle=90]{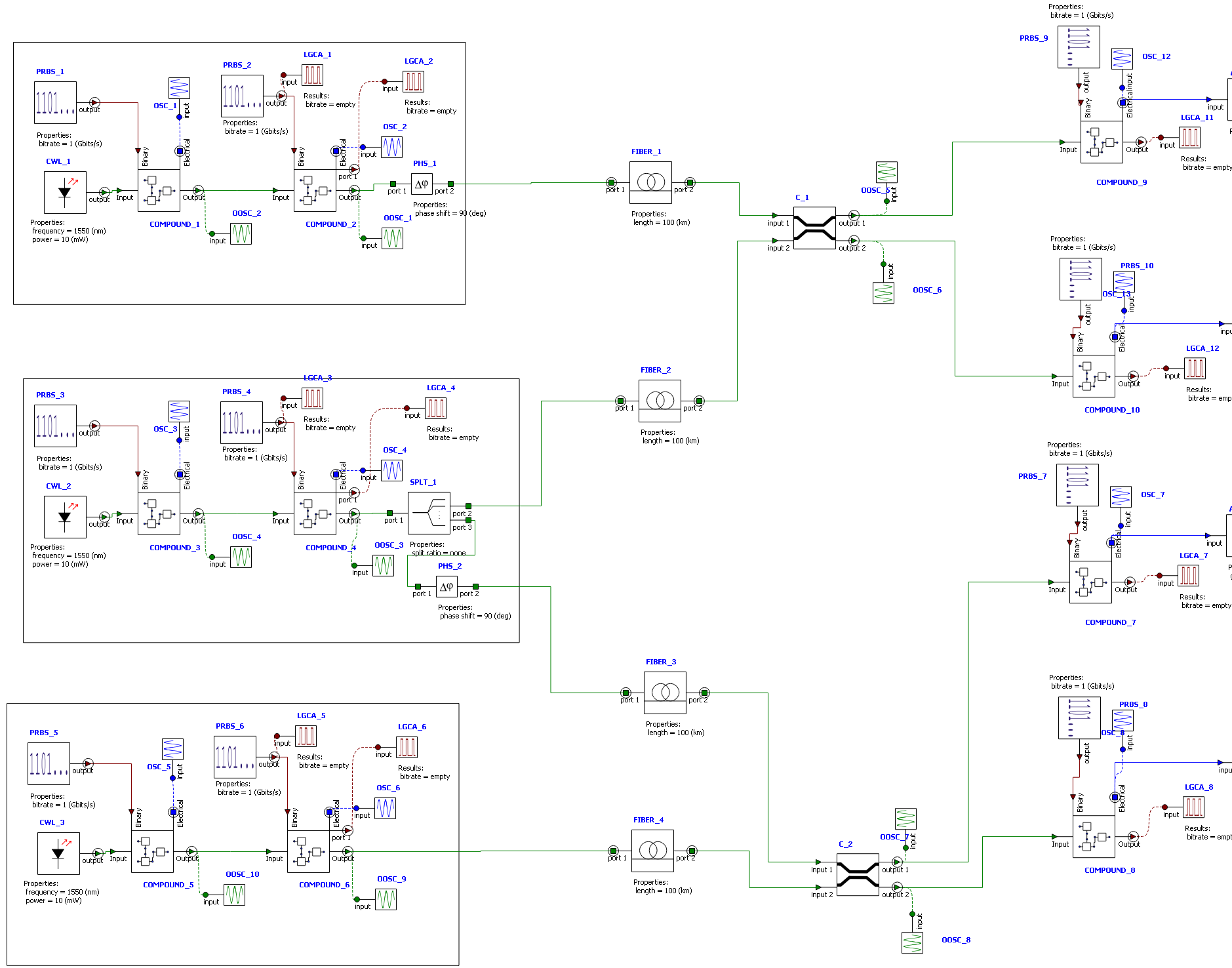}
    \caption{ANSYS Interconnect simulation for multi-party twin field protocol.}
    \label{ansys}
\end{figure}
\color{black}
Using simulations provides a convenient and powerful way to comprehend the workings of the QKD systems and the protocol implemented, when a full analytic treatment proves to be elusive. Only the key rate of the idealized protocol, which involves the quantum states based on attacks described in Sec. 2.2, is captured by our preceding theoretical analysis of security. The complete QKD security evaluation must consider not only the protocol scheme but also practical device flaws. Carefully evaluating the functional QKD systems and their defects requires taking into account the quantum states of the transmitted photons as well as the optical components. Simulations can be used to examine the theory-practice discrepancies and study countermeasures. The operational characteristics of the optical equipment, as well as environmental or Eve's interference, are necessary to study a complete simulation. Initial QKD simulations used ideal assumptions and few optical components to numerically measure the performance \cite{Shuang2008,Coles2016}. For the polarization-based QKD protocols such as BB84, extensive packages have been made available \cite{Buhari2012,mailloux2015,PhysRevApplied.14.024036,Fan-Yuan2020}. Additionally, while polarization-based QKD methods are subjected to hacking attempts such as photon-number-splitting \cite{mailloux2016}, hacking on measurement device-independent protocols like twin-field is different \cite{PhysRevApplied.13.034008,app11209560,IAQKD,peng2024practicalsecuritytwinfieldquantum}. We perform simulations of the protocol on the ANSYS Interconnect platform for finite key rate effects and describe the practical implementation with optical components. 
\color{black}

The image in Fig. \ref{ansys} shows a simulation for the multi-party twin-field protocol defined in Sec. 2.1. The rectangular selections show the modules of Alice, Bob and Charlie from top to bottom on the left side of Fig. \ref{ansys}. 

Considering the Alice module, single-mode continuous wave (CWL\_1) laser element from the ANSYS Interconnect library is used to model a continuous optical source. The laser produces an optical signal with a single frequency and constant amplitude. The laser's phase is specified and remains steady throughout the simulation.

\begin{figure}[htp]
    \centering
\includegraphics[width=17cm]{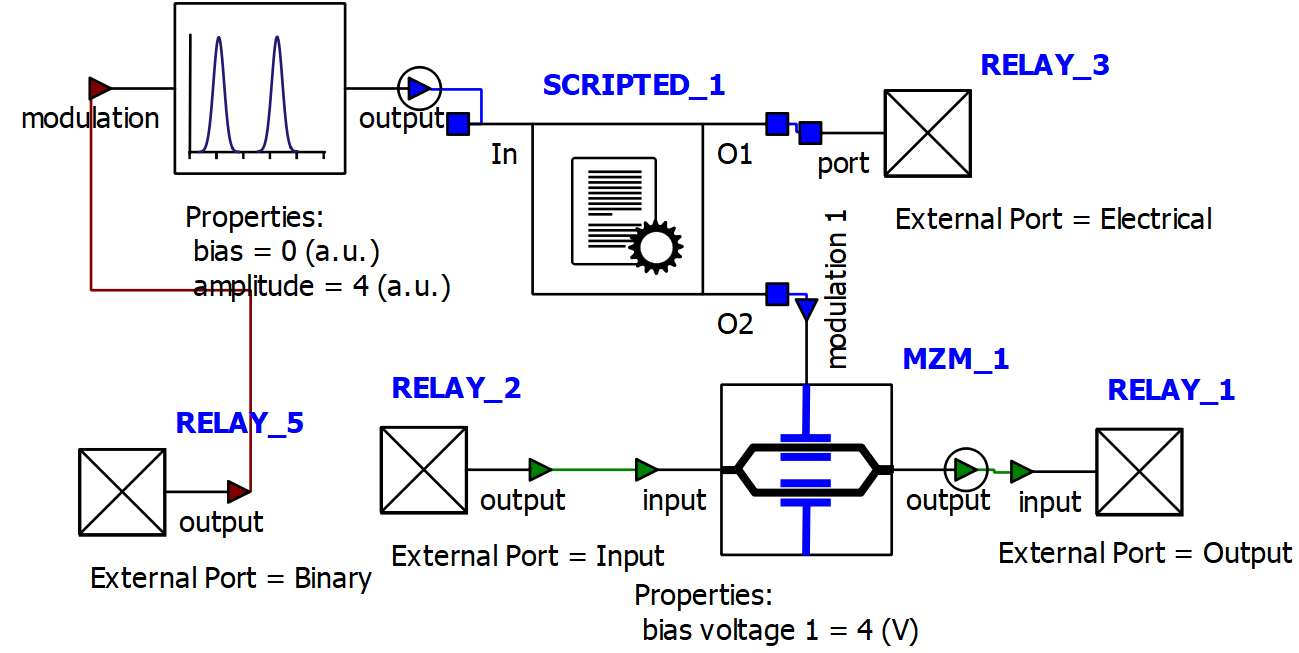}
    \caption{ANSYS Interconnect simulation for intensity modulator in Alice, Bob and Charlie modules.}
    \label{IM}
\end{figure}

Further, a train of optical pulses are produced from the optical signal using an intensity modulator (COMPOUND\_1, A compound element on ANSYS). Building complicated photonic integrated circuits based on hierarchical sub-circuit elements requires compound elements on ANSYS. COMPOUND\_1 is displayed in Fig. \ref{IM}. To model the intensity and phase modulators, a Mach-Zehnder modulator (MZM\_1) component from the ANSYS Interconnect library is utilised. The optical input ($E_{in}$) of the MZM is divided into the upper and lower arms. Phase shifts $\phi_1$ and $\phi_2$ are then used to phase modulate the optical input, with the electrical signals $V_1$ and $V_2$ driving them. Phase shifts $\phi_1$ and $\phi_2$ are given as \begin{equation}
\phi_{1(2)}=\pi\left(\frac{V_{1(2)}}{V^{\pi}_{\text{RF}}}+\frac{V^{\text{bias}}_{1(2)}}{V^{\pi}_{\text{DC}}}\right),
\end{equation} and the two modulated signals are then recombined to yield optical output $E_{out}$, defined by \begin{equation}
    E_{out} = E_{in} \left(\text{e}^{i\phi_1}+\text{e}^{i\phi_2}\right).
\end{equation}

The input radio frequency (RF) voltage and bias voltage applied to two arms with RF voltage $V^{\pi}_{\text{RF}}$ and DC voltage $V^{\pi}_{\text{DC}}$ are denoted by $V_{1(2)}$ and $V_{1(2)}^{\text{bias}}$. The RF voltage from the pulse generator is utilised to modulate the optical signal input from the laser source. The pulse generator powered by pseudo-random binary sequence (PRBS) generates periodic electrical pulses. When the MZM (MZM\_1 in Fig. \ref{IM}) is in the "balanced single drive" mode, the two arms get equal and opposite voltages from the RF pulse generator by setting the PRBS\_1 (Input to intensity modulator COMPOUND\_1 in Alice module of Fig. \ref{ansys}) output to "1"s. With the appropriate bias voltage, the laser source is converted into a pulsed signal.  

\begin{figure}[htp]
    \centering
\includegraphics[width=17cm]{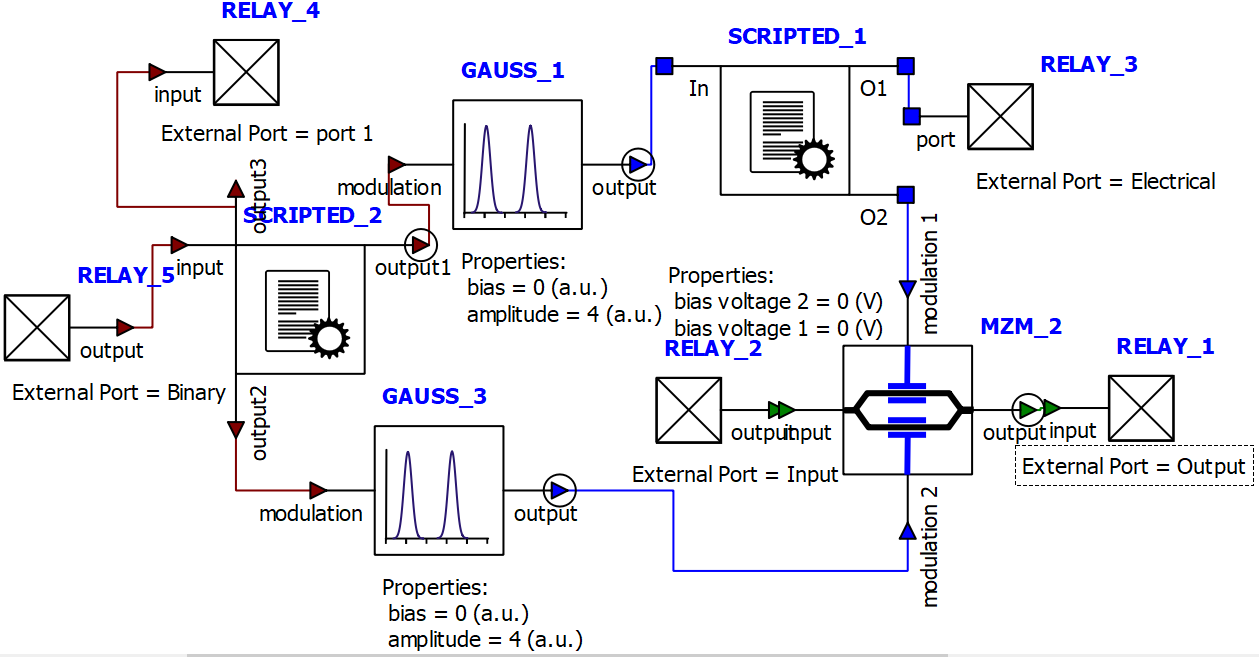}
    \caption{ANSYS Interconnect simulation for phase modulator in Alice, Bob and Charlie modules.}
    \label{PM}
\end{figure}

Further, a phase modulator (COMPOUND\_2, as seen in Fig. \ref{PM}) processes the optical pulses from the intensity modulator. An MZM (MZM\_2) is used to realise the phase modulator, and the same RF voltage is delivered to both arms. As a result, the RF pulse generator's input $k_A = 0(1)$ for PRBS\_2 (Input to phase modulator COMPOUND\_2 in Alice module of Fig. \ref{ansys}) produces optical pulses whose phases are modulated by $0(\pi)$. 

\begin{figure}[htp]
    \centering
\includegraphics[width=17cm]{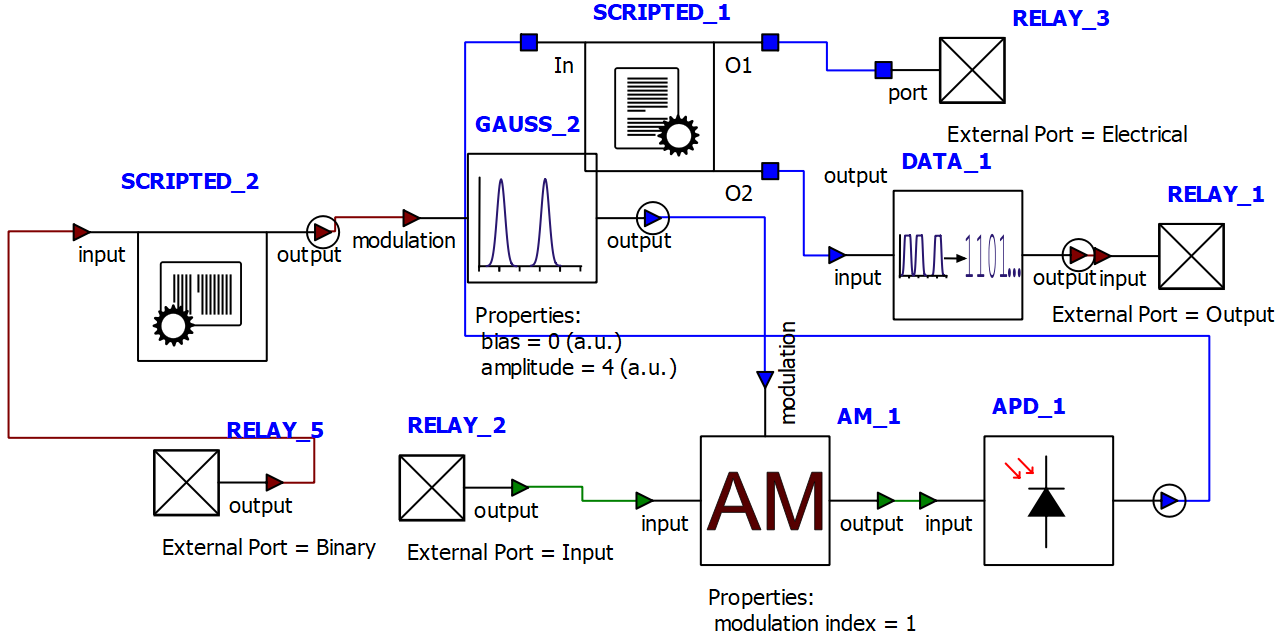}
    \caption{ANSYS Interconnect simulation for detectors in nodes AB and BC.}
    \label{SPD}
\end{figure}

Similarly, Bob and Charlie modules produce these laser pulses, which are phase-modulated randomly by either $0$ or $\pi$ defined by $k_B$ and $k_C$. Alice sends it through the quantum channels FIBER\_1; Bob sends the pulses by keeping a splitter (SPLT\_1) where the same optical input will go through both output ports FIBER\_2 and FIBER\_3 and Charlie sends it through FIBER\_4. When two pulses arrive at the intermediate nodes AB(BC) consisting of $2\times2$ coupler ($50:50$ coupling ratio)  if the coupler C\_1(C\_2) finds a pair of pulses with same phase, it is fed to photodetector COMPOUND\_9 (COMPOUND\_7); if it finds pulses with different phase, it is fed to photodetector COMPOUND\_10 (COMPOUND\_8). The Avalanche photodetector (APD\_1) element of ANSYS Lumerical Interconnect was utilised in our simulation as can be seen in COMPOUND\_9 displayed in Fig. \ref{SPD}. The probability $Q_\mu$ of a detection event when the pulse reaches APD\_1 is given as \cite{PhysRevA.72.012326} \begin{equation}
    Q_\mu = Y_0 + 1-\text{e}^{-\eta\mu}
\end{equation} where $\mu$ is the mean photon number of the pulse. The background contributions of noise in the quantum channel are taken into account by $Y_0$. Since the quantum channel and classical communication occur on the same fibre, Raman noise also plays a role in $Y_0$. With the dark count probability of $1\times10^{-6}$ per clock cycle of the employed detector, $Y_0$ is considered $2.45\times10^{-6}$ \cite{Wang2017}. The transmittance is calculated as $\eta=10^{-0.2L/10}$ for distance $L$ in km.

\begin{figure}[htp]
\centering
\begin{subfigure}{0.45\linewidth}
\includegraphics[width=\linewidth]{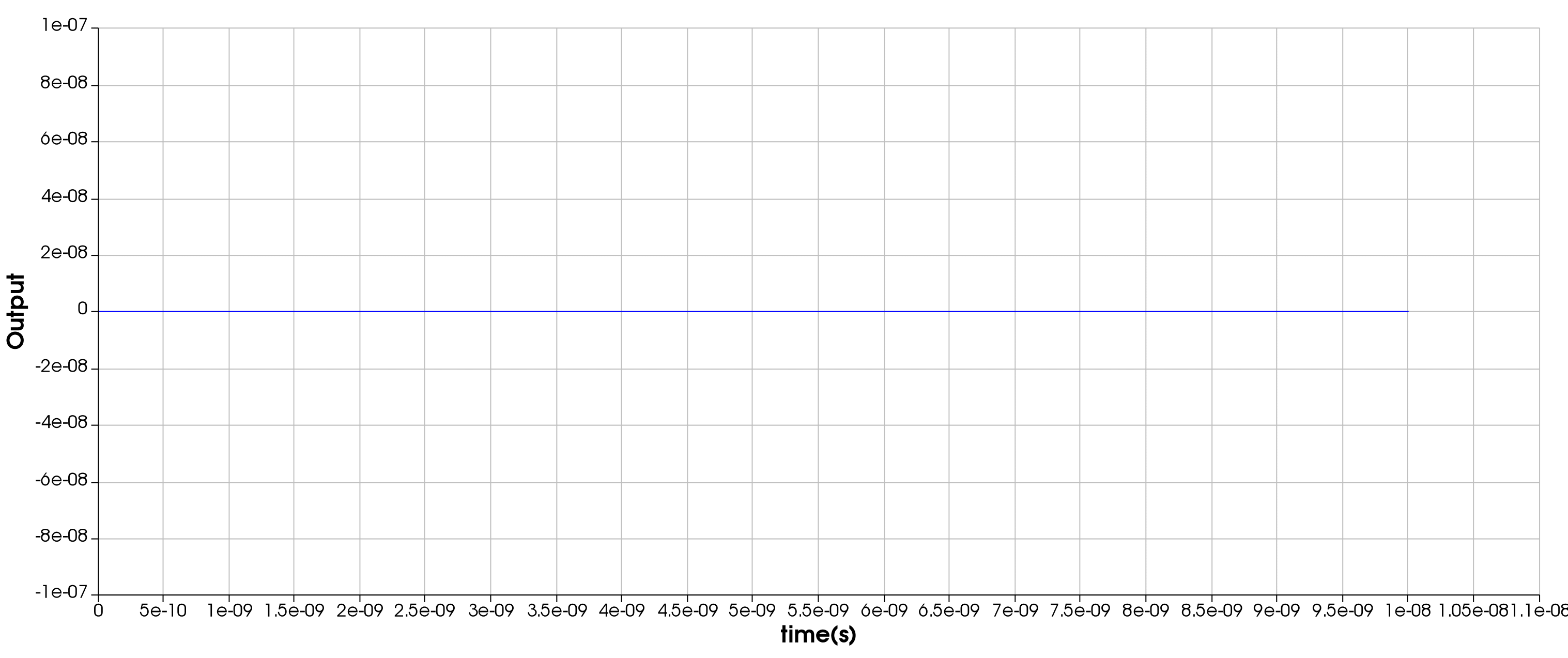}
\caption{PRBS\_2 output measured in LGCA\_1 (Alice Module)}
\end{subfigure}
\begin{subfigure}{0.45\linewidth}
\includegraphics[width=\linewidth]{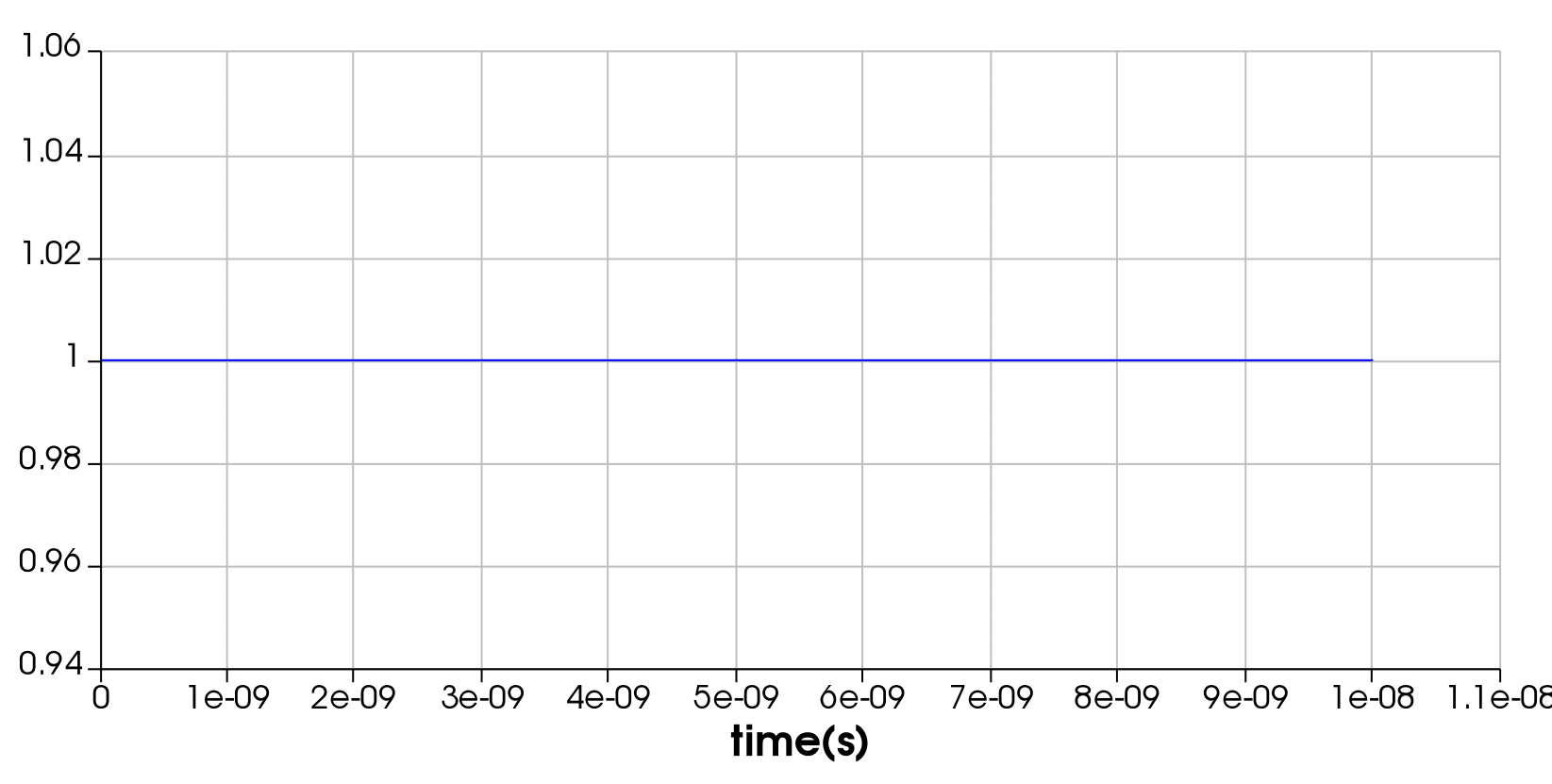}
\caption{PRBS\_4 output measured in LGCA\_3 (Charlie Module)}
\end{subfigure}
\begin{subfigure}{0.45\linewidth}
\includegraphics[width=\linewidth]{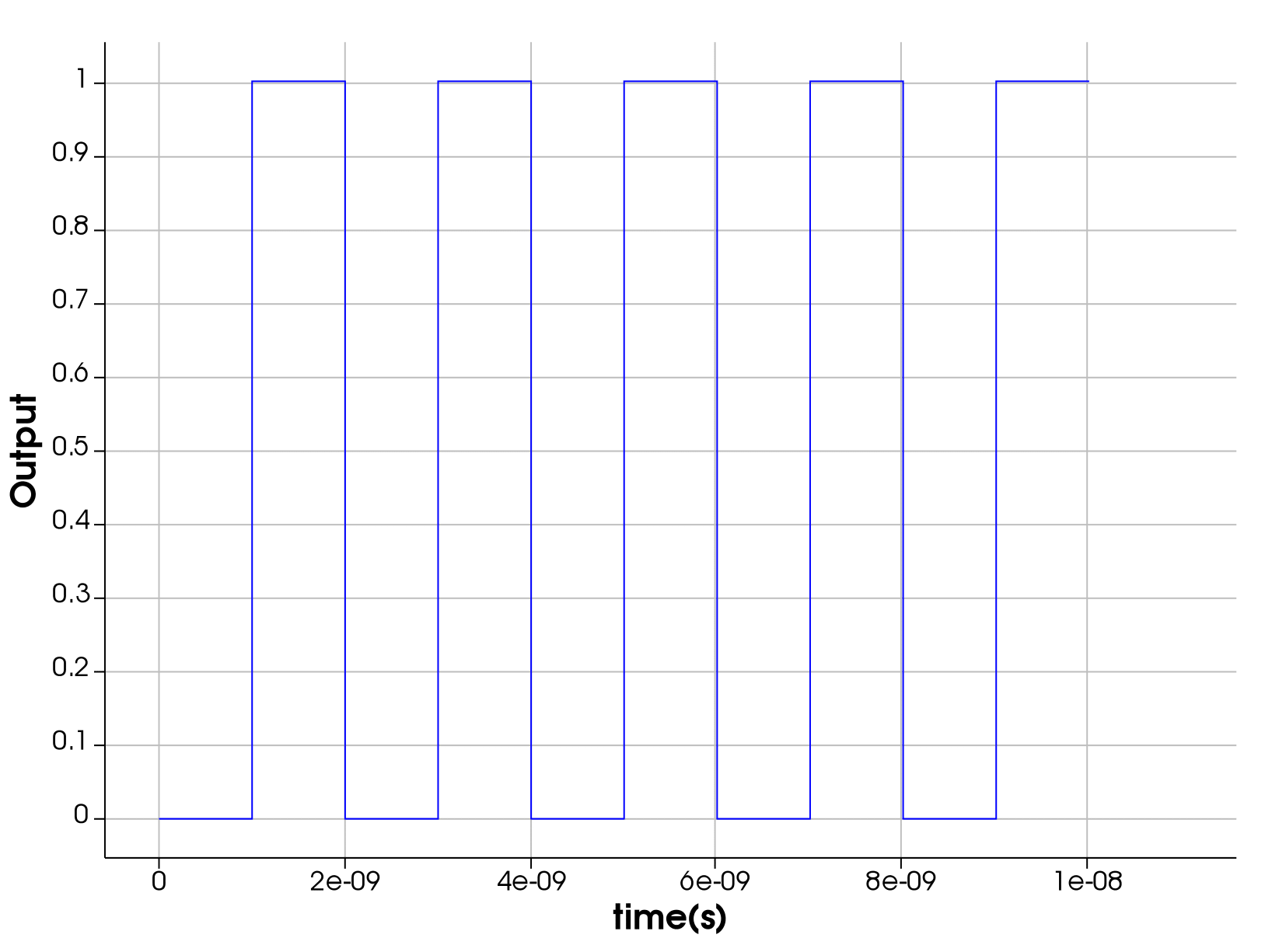}
\caption{PRBS\_6 output measured in LGCA\_5 (Bob Module)}
\end{subfigure}
\caption{Comparing phase modulator output of Alice, Bob and Charlie modules from PRBS\_2, PRBS\_4 and PRBS\_6, respectively in digital form.}
\label{PMpulses}
\end{figure}

\begin{figure}[htp]
\centering
\begin{subfigure}{0.45\linewidth}
\includegraphics[width=\linewidth]{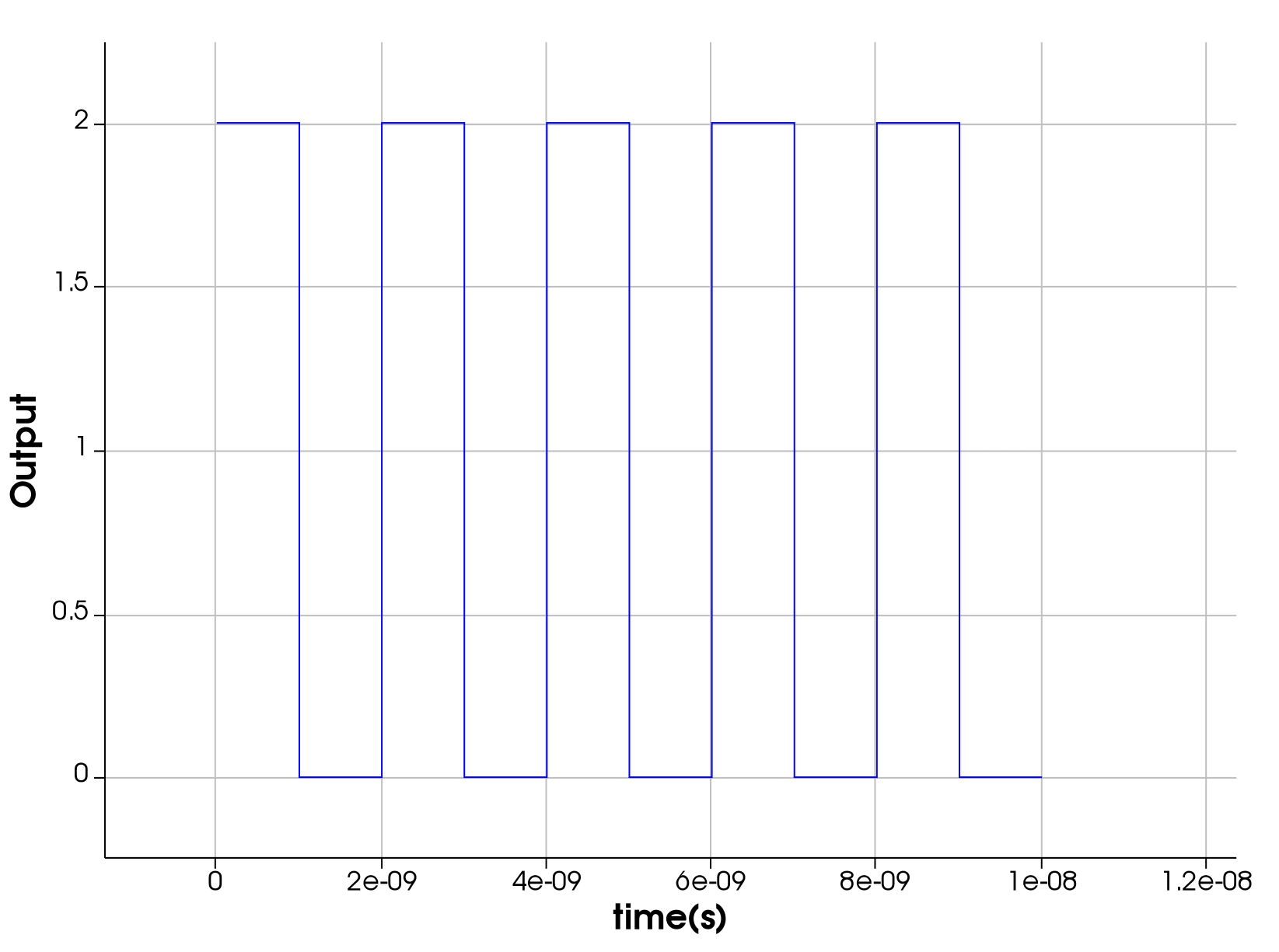}
\caption{Detector COMPOUND\_9 output measured in LGCA\_11 \\ (Node AB)}
\end{subfigure}
\begin{subfigure}{0.45\linewidth}
\includegraphics[width=\linewidth]{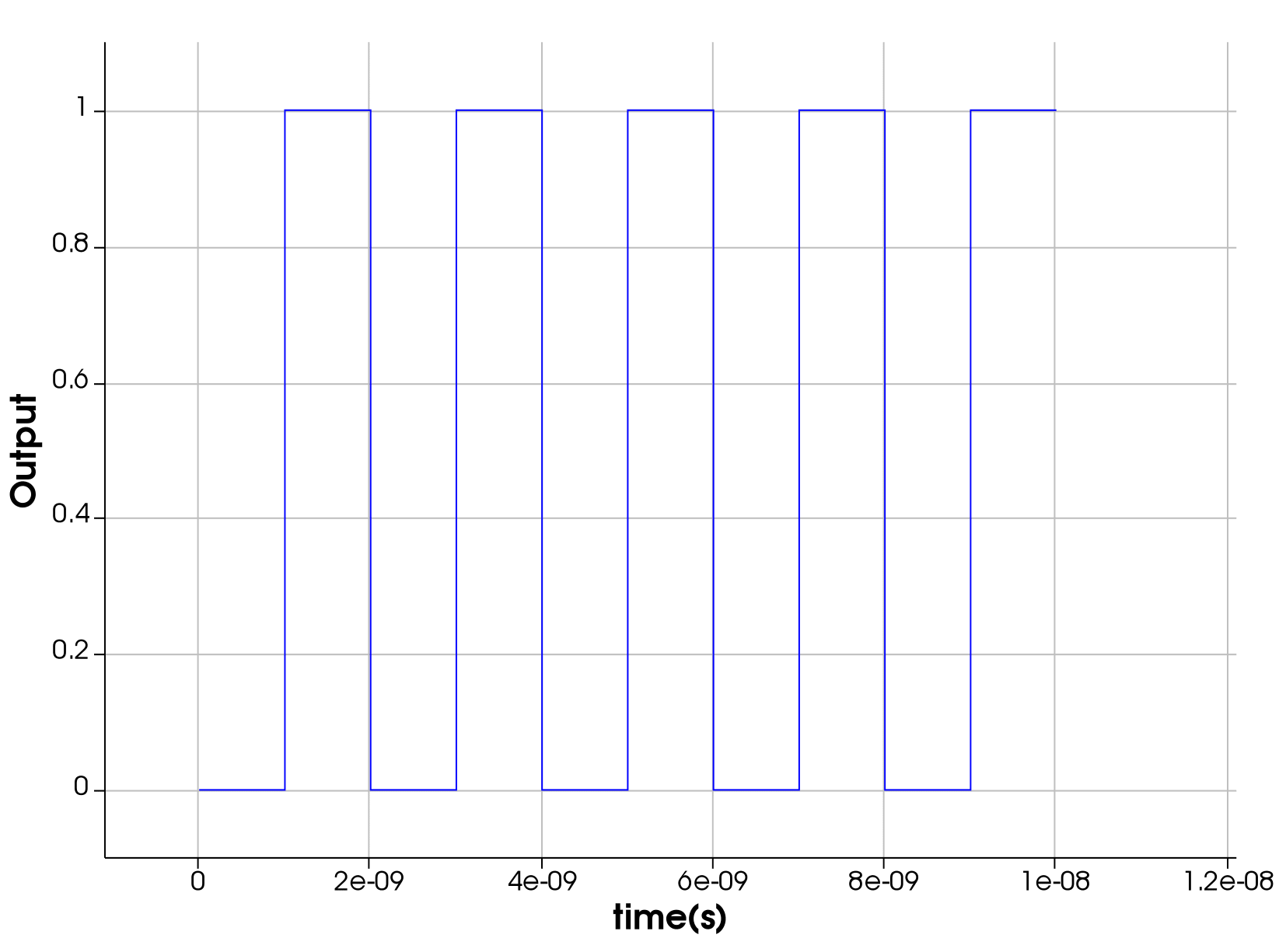}
\caption{Detector COMPOUND\_10 output measured in LGCA\_12 (Node AB)}
\end{subfigure}
\begin{subfigure}{0.45\linewidth}
\includegraphics[width=\linewidth]{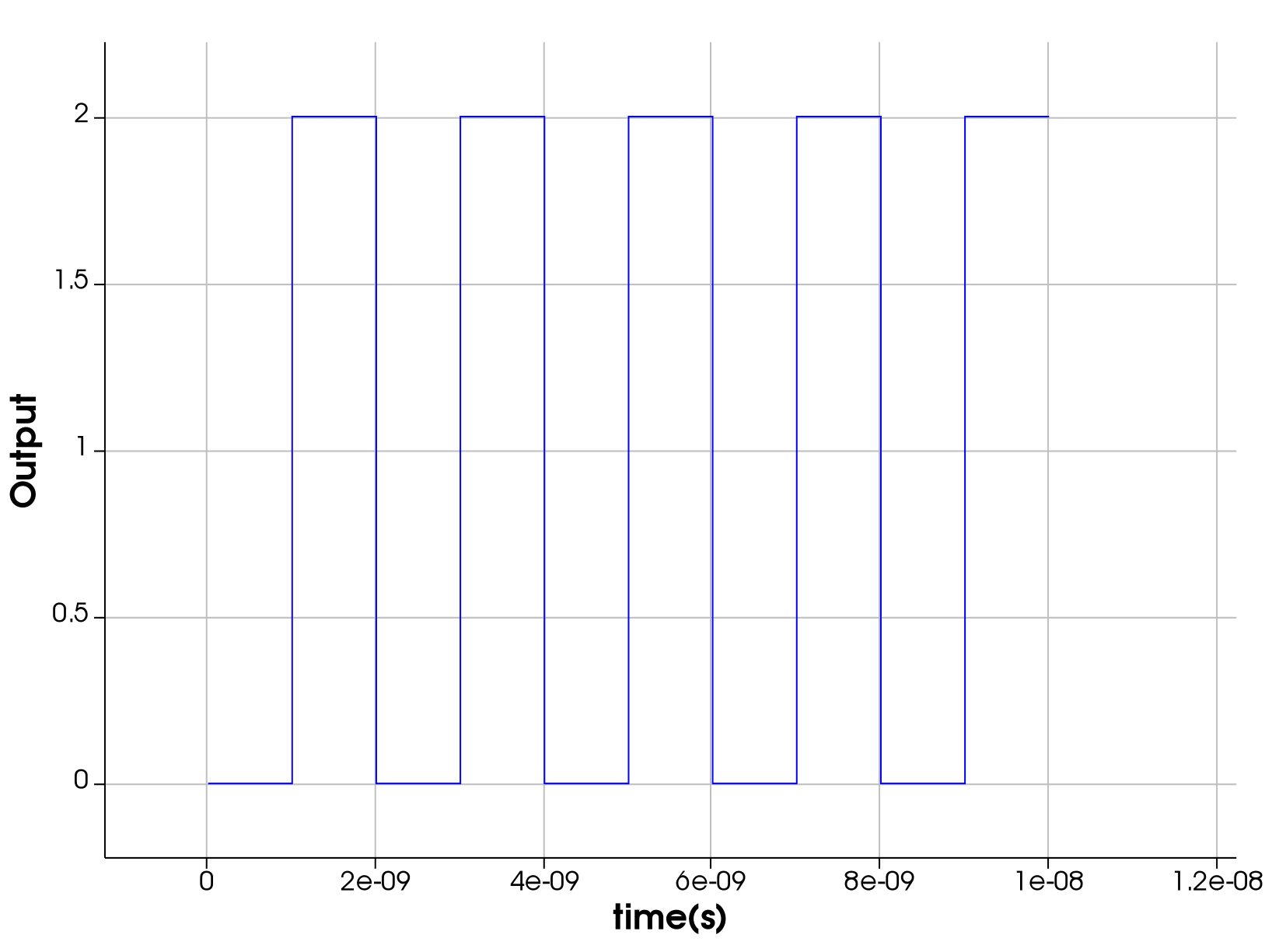}
\caption{Detector COMPOUND\_7 output measured in LGCA\_7 (Node BC)}
\end{subfigure}
\begin{subfigure}{0.45\linewidth}
\includegraphics[width=\linewidth]{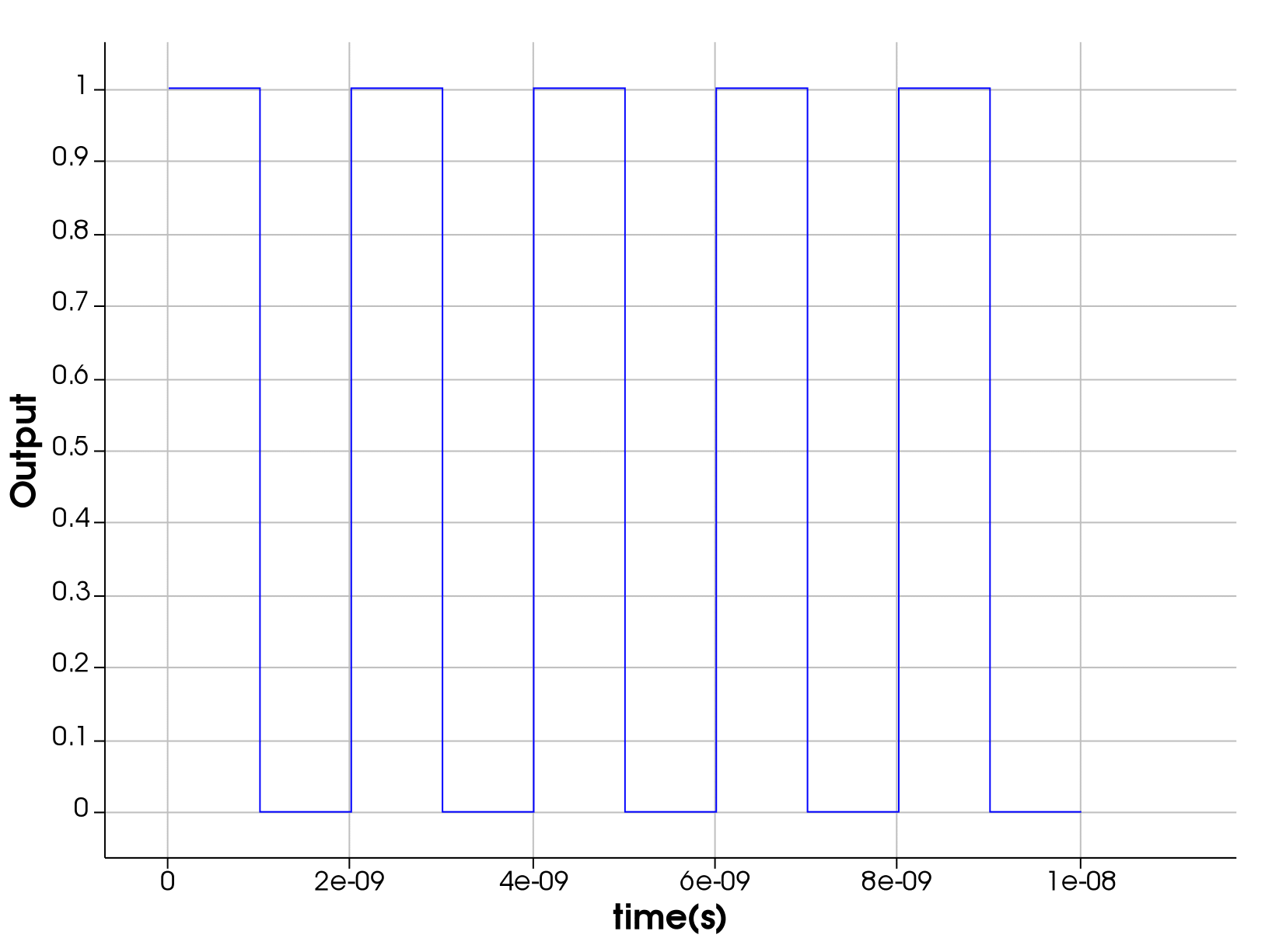}
\caption{Detector COMPOUND\_8 output measured in LGCA\_8 (Node BC)}
\end{subfigure}
\caption{Comparing detections at node AB from C\_1 detectors COMPOUND\_9, COMPOUND\_10  and node BC from C\_2 detectors COMPOUND\_7, COMPOUND\_8, respectively in digital form.}
\label{nodepulses}
\end{figure}

For illustrating a particular instance without losses, we consider Alice sending pulses with 0 phase (PRBS\_2 with output 0), Bob sending alternate pulses with 0 and $\pi$ phase (PRBS\_4 with output 0 and 1 alternately), Charlie sending pulses with $\pi$ phase (PRBS\_6 with output 1) as shown in Fig. \ref{PMpulses}. Based on this input the output measured in detectors at nodes AB (COMPOUND\_9 and COMPOUND\_10) and BC (COMPOUND\_7 and COMPOUND\_8) is displayed in Fig. \ref{nodepulses}. In our case amplitude zero indicates the no detection, amplitude one indicates detection with different phase, when fed to photodetector COMPOUND\_10(COMPOUND\_8) and amplitude two indicates detection with same phase, when fed to photodetector COMPOUND\_9(COMPOUND\_7) at the node AB(BC) with coupler C\_1(C\_2).

\begin{figure}[htp]
    \centering
\includegraphics[width=13cm]{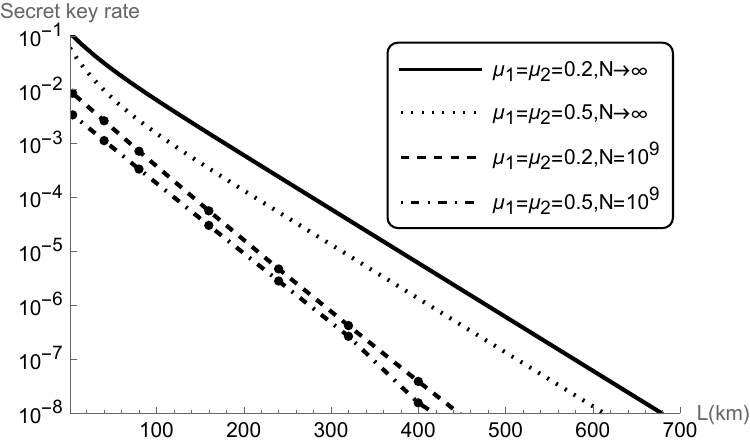}
    \caption{\color{black}Secret key rate from simulations compared with asymptotic key rate from Eq. (26). The optimal value of mean photon number $\mu$ for higher secret key rate is around 0.2 with finite pulses.}
    \label{keyrate}
\end{figure}

\color{black}
We consider 1 GHz pulse repetition rate with the line width of the laser as 5 kHz and analyze the key rate by performing multiple iterations of the simulations for varied distances $L$ in km with losses described by Eq. (29) ($L=l_A+l_B+l'_B+l_C$ is fixed where $l_A=l_B=l_B'=l_C$ in FIBER\_1, FIBER\_2, FIBER\_3, FIBER\_4 of Fig. \ref{ansys}). We obtain the secret key rate in the log scale, which closely matches the behavior of the result from the previous section, as shown in Fig. \ref{keyrate}. The final secret key rate (skr) is obtained from the sifted key rate (kr) based on the Holevo bound from Eq. (19), where $\hbox{skr}=\hbox{kr}\times(1-\chi)$. For a finite number of pulses sent ($N=10^9$), the key rates are lower than the asymptotic key rate from Eq. (26), as expected. Further, in Fig. \ref{keyrate}, we observe that by varying the mean photon number, $\mu=0.2$ shows optimal secret key rate compared to $\mu=0.5$ (considering $\mu_1=\mu_2=\mu$ in Eq. (26)). The secret key rate was more than $10^{-6}$ per pulse (1 kbps) for $L=$ 250kms. 
\color{black}
\section{Difficulty in practical implementation}

Practically putting in place a twin-field protocol between multiple parties presents technological difficulties since active phase stabilization is necessary to interfere with coherent pulses due to random phase fluctuations \cite{Minder2019,PhysRevLett.123.100505,Fang2020,PhysRevLett.124.070501}. \color{black} It is worth noting that this problem presents the main bottleneck in realizing twin-field cryptography, and calls for unique technology developed to address it. Here we mention two such methods: (a) in one case, the wavelengths of two users' operating two separate lasers are locked to estimate and compensate for phase drift \cite{PhysRevLett.123.100505, PhysRevLett.124.070501}; (b) In another case, phase fluctuation was overcome by matching the modes of two independent lasers using the laser injection technique and the phase post-compensation approach \cite{Fang2020}. \color{black} The central nodes AB and BC function as gigantic Mach-Zehnder interferometers for interference measurement, as shown in Fig. \ref{multiparty}. Stabilizing an unbalanced Mach-Zehnder interferometer with widely different path lengths is challenging. So, the setup would necessitate that users Alice-Bob (Bob-Charlie) have an identical geographic distance to the central node AB(BC) for the interference of pulses ideally, whereas, in the real world, there could be significant variations in the geographic distances between various parties and the nodes. A more practical method to accomplish this is via sending-or-not-sending, a unique type of twin-field protocol where data is encoded based on the decision to send in a coherent state (sending) or a vacuum state (not sending), and it has tolerance to phase misalignment errors \cite{PhysRevA.98.062323}. 
 
 Following the discussion similar to in Sec. 2 using the setup in Fig. \ref{multiparty}, Alice, Bob and Charlie decide at random whether to send a vacuum state with probability $1-p_s$ recording a bit value 0, 1 and 0 or a coherent state $\ket{\alpha}$, $\ket{\beta}$ and $\ket{\gamma}$ with probability $p_s$ recording a bit value 1, 0 and 1. They get a bit error when both Alice and Bob (Bob and Charlie) decide to send the coherent state, and the click is recorded in node AB (BC). They must select a small sending probability $p_s$ to lower the likelihood of a sending-sending case $p^2_s$ since the correct clicks correspond to the case where only one of Alice and Bob (Bob and Charlie) choose to send a coherent state to node AB (BC). It was shown recently that actively odd-parity pairing post-processing technique can address this issue by enhancing the transmission probability and notably lower the impact of bit errors \cite{PhysRevLett.130.210801}. 

\section{Conclusion}
We propose and examine a technique that expands on the two-party twin-field quantum key distribution to obtain a protocol for the concurrent multi-party quantum key agreement. For the proposed protocol, we carry out the asymptotic key rate analysis from a security proof based on the technique of entanglement-based source replacement. We simulated the protocol on the ANSYS Interconnect platform by assessing its practical implementation. These simulations can be augmented by including realistic experimental issues such as phase misalignment errors and the use of decoy states with varying intensities. This exercise can be extended to extensively study the variant of the present protocol based on the sending-or-not-sending twin-field protocol \cite{PhysRevA.98.062323}. These simulations can also be used to quantify the information lost by the quantum hacking efforts \cite{RevModPhys.81.1301,RevModPhys.92.025002,IAQKD}. \color{black} Trojan-horse attack \cite{jmo2001,th2006,Jain_2014,Sajeed2017,Jain2015} and backflash attack \cite{Pinheiro:18,Meda2017} were studied on distributed-phase-reference QKD simulations using ANSYS, which might also be relevant to twin-field implementation \cite{abhignan2024simulations}. \color{black}
 
\bibliographystyle{ieeetr}
\bibliography{sample.bib}
\end{document}